\documentclass[preprint,nofootinbib,prd,aps]{revtex4}
\usepackage{graphicx,amstext,amssymb,amsmath,color}
\begin{document}

\title{Cosmological particle creation in the little bang}

\author{S.V. Akkelin$^{1}$}
\affiliation{$^1$Bogolyubov Institute for Theoretical Physics,
Metrolohichna  14b, 03143 Kyiv,  Ukraine}

\begin{abstract}

Particle production  by expanding in the future light cone scalar quantum field 
is studied by assuming that the 
initial state  is associated with  the   quasiequilibrium
statistical operator corresponding to fluid dynamics.  We calculate  particle production from a 
longitudinally boost-invariant expanding quantum field designed as a simple but reliable model
for the central rapidity region of a relativistic  collision.  Exact diagonalization of 
the model is performed  by introducing a
notion of quasiparticles. 

\end{abstract}

%\pacs{}

 \maketitle

 \section{Introduction}

It is noteworthy that   relativistic hydrodynamics can be successfully 
applied for a phenomenological description of particle production in
the relativistic nucleus and particle collisions; for a recent review
see, e.g., Ref. \cite{Shen}.   Applicability  of hydrodynamics 
for calculations of   particle momentum spectra    indicates  
the validity of some  reduced descriptions for the state of a system.
Such a  reduced description can be fulfilled (and generalized hydrodynamic equations  
can be   derived)  utilizing  Zubarev's formalism of a
nonequilibrium statistical operator; see, e.g., Refs. \cite{Zubarev-1,Zubarev-2,Zubarev-3,Zubarev-4}   
and  also  Refs. \cite{Bec-1,Bec-2,Bec-3,Bec-4,Bec-5} for recent papers related to this method.  
A basic component of this formalism is the  quasiequilibrium statistical operator.
The quasiequilibrium statistical operator
$\rho_{\sigma}^{q}$ is constructed by maximizing  information
entropy \cite{Janes-1,Janes-2} for the average local values of the
energy-momentum density  operator on a three-dimensional spacelike hypersurface $\sigma_{\mu}$ with a
timelike normal vector $n_{\mu}(x)$:
\begin{eqnarray}
n_{\mu} (x) \langle T^{\mu \nu}_{ren}(x)\rangle = n_{\mu} (x) \langle T^{\mu \nu}_{ren}(x)\rangle_{\sigma}^{q}. \label{1.2}
\end{eqnarray}
The left-hand side in the above equation  is  calculated with a true state of the system,
and the right-hand side is calculated
with the quasiequilibrium statistical 
operator, $\langle ...\rangle_{ \sigma}^{q} \equiv Tr[\rho_{\sigma}^{q}...]$. Here  $T^{\mu \nu}_{ren}(x)$ denotes 
the renormalized  operator of an energy-momentum tensor:  It is well known  that expectation values of
an energy-momentum tensor need regularization  because   expectation values of the products 
of quantum field operators taken in 
the same spacetime point are ill defined in the general case. 
The  quasiequilibrium statistical operator reads 
\begin{eqnarray}
\rho_{\sigma}^{q}=Z^{-1}_{\sigma} \exp\left(-\int_{\sigma } d\sigma n_{\mu}(x)\frac{u_{\nu}(x)
T^{\mu \nu}(x)}{T(x)}\right ), \label{1}
\end{eqnarray}
where $T(x)$ is the local  temperature,
$u_{\nu}(x)$ is hydrodynamical four-velocity,
$u_{\nu}(x)u^{\nu}(x)=1$, 
$\sigma_{\mu}$ is a three-dimensional spacelike hypersurface with a
timelike normal vector $n_{\mu}(x)$, $T^{\mu \nu}(x)$ is the
operator of an energy-momentum tensor, and $Z_{\sigma}$ is the normalization
factor making $Tr[\rho_{\sigma}^{q}]=1$. We assume for simplicity that a chemical
potential $\mu = 0$. Equation (\ref{1.2}) defines the  $u^{\nu}(x)/T(x)$ field in Eq. (\ref{1}).
It is worth noting  
that the  quasiequilibrium statistical
operator (\ref{1}) can have a parametric time dependence even in the  Heisenberg representation.

The zero temperature
limit of such a   
statistical operator  defines the corresponding ground state. If one has $n^{\mu}(x)=u^{\nu}(x)=(1,\textbf{0})$,
and  $T =
\mbox{const}$, then $ \int_{\sigma }d\sigma
(n_{\mu}(x)u_{\nu}(x)T^{\mu \nu}(x)/T(x))=H/T$ where $H$ is the Hamiltonian. In this case the 
 ground state is $| 0 \rangle \langle 0|$, where $| 0 \rangle$ is  the eigenstate of the
 Hamiltonian with the lowest eigenvalue and, therefore, coincides with the ordinary Minkowski vacuum.
However, in general, $ \int_{\sigma }d\sigma
(n_{\mu}(x)u_{\nu}(x)T^{\mu \nu}(x)/T(x)) \neq H/T$, and therefore the  ground state
(``vacuum'') of the quasiequilibrium statistical operator (\ref{1})
may  be different from the ordinary  Minkowski vacuum. 
Perhaps  the most
straightforward definition of such a  ground state can be performed 
by means of the  instantaneous diagonalization of the operator  $ \int_{\sigma }d\sigma
(n_{\mu}(x)u_{\nu}(x)T^{\mu \nu}(x)/T(x))$ in some quasiparticle
basis. The instantaneous ground state,  $| 0_{ \sigma } \rangle$, can  be defined as the
eigenstate of this normal ordered  operator with the lowest eigenvalue
at  the three-dimensional  hypersurface $\sigma_{\mu}$.  Of course, the  existence of such an 
operator with a non-negative spectrum, bounded from below, is crucial to the 
existence and determination of the  ground state containing
no quasiparticle  excitations.  

The validity of the reduced   description means that the  quasiequilibrium statistical operator
(\ref{1}) can be  utilized for (approximate) calculations of various operator-valued quantities, 
i.e.,  it can be 
regarded as a reliable proxy for a true state of a system. Typically, 
the expansion of an isolated system   leads  to a breakdown  of the reduced description. 
Calculations at this far-from-equilibrium stage can be essentially simplified  for the extreme case
of a sudden   decoupling  of the quasiequilibrium state of a system which can be  conditioned by switching off the 
interactions due to expansion.
In the present paper we  analyze particle creation after the  sudden   decoupling  of a system  utilizing  
simple but phenomenologically relevant  model of  
the quasiequilibrium state for   a boost-invariant expanding noninteracting  boson quantum field. 
Unlike in previous studies  \cite{Sinyukov-1,Sinyukov-2,Sinyukov-3,Akkelin}, 
we  perform exact  diagonalization of the model in a quasiparticle basis. 
We show that the ground state of such a model is time dependent and does not coincide with the Minkowski vacuum.
Then we estimate momentum 
spectra of particles produced as a  result of the sudden   decoupling  of the quasiequilibrium state. 

\section{Diagonalization of the quasiequilibrium  statistical operator for  boost-invariant expanding quantum  scalar field}

In this section we diagonalize  quasiequilibrium statistitical operator  (\ref{1}) for the 
longitudinally  boost-invariant 
Bjorken-type \cite{Bjorken} expanding   scalar field. This type of expansion is 
relevant for an ultrarelativistic collision and allows us to perform  calculations analytically.
Namely,  we assume  that matter produced in the  little bang  is locally restricted to
the light cone beginning at the  $t=z=0$ plane of the Minkowski
spacetime manifold. Because the spacetime region occupied by the matter produced 
in an ultrarelativistic  collision lies in the future light cone of a  collision,
it is convenient to introduce the   Bjorken coordinates $(\tau , \eta)$
instead of the Cartesian ones $(t,z)$: 
\begin{eqnarray}
t = \tau \cosh \eta, \label{2} \\
z = \tau \sinh  \eta ,  \label{3} 
\end{eqnarray}
where $\eta$ is the so called  spatial rapidity, $\tanh\eta = z/t $.
The two other coordinates $\textbf{r}_{T}=(r_{x},r_{y})$ are the
Cartesian ones. 
One can see that $(\tau , \eta)$ coordinates cover the whole future light cone region.
The Minkowski line element restricted to the light
cone  has then the form  [we use the convention $g^{\mu \nu}=\mbox{diag}(+1,-1,-1,-1)$]:
\begin{eqnarray}
ds^{2}=dt^{2}-d\textbf{r}_{T}^{2}-dz^{2}=  d\tau^{2}- d\textbf{r}_{T}^{2} - \tau^{2}d
\eta^{2} . \label{4}
\end{eqnarray}
Notice that  $\tau$ is the proper time of observers
which move with different but constant longitudinal velocities in
such a way that their world lines begin at $z=t=0$. The corresponding longitudinally 
boost-invariant  four-velocity $u^{\mu}$ is expressed as  
\begin{eqnarray}
u^{\mu}(x)=(\cosh \eta , 0,0, \sinh  \eta). \label{5}
\end{eqnarray}
One can see  that $\tau$ also controls a value of the  four-velocity spacetime gradients,
\begin{eqnarray}
\partial_{\mu}u^{\mu}(x)=\frac{1}{\tau}. \label{5.1}
\end{eqnarray}

We assume that  a quasiequilibrium state is  defined at a hypersurface with constant energy
density in the comoving coordinate system for the  field expanding with the four-velocity 
(\ref{5}). Then   $T (x)$ is constant on the
corresponding hypersurface, and such a  hypersurface is defined by
constant $\tau = \sqrt{t^{2}-z^{2}}$. This implies that
\begin{eqnarray}
n^{\mu}(x)=u^{\mu}(x), \label{6} \\
T (x)= T(\tau), \label{6.1}
\end{eqnarray}
and
\begin{eqnarray}
d \sigma  = \tau d\eta dr_{x}dr_{y}.  \label{7}
\end{eqnarray}
Then the  quasiequilibrium statistical operator (\ref{1})
reads
\begin{eqnarray}
\rho_{\tau}^{q}=Z^{-1}_{\tau} \exp\left(- \frac{H^{[\tau]}}{T(\tau)}\right ), \label{13}
\end{eqnarray}
where 
\begin{eqnarray}
H^{[\tau]}= \int  d\sigma u_{\mu}(x)u_{\nu}(x)T^{\mu \nu}(x) . \label{14}
\end{eqnarray}

Here we are primarily interested in the effects of
particle creation due to the expansion of quantum fields.  Therefore
we  disregard  field self-interactions and
consider  a  scalar quantum field model with the classical action 
\begin{eqnarray}
S=\int dt d^{3}r \left [\frac{1}{2} \left (\frac{\partial
\phi}{\partial t}\right )^{2} - \frac{1}{2} \left (\frac{\partial
\phi}{\partial \textbf{r}}\right )^{2} - \frac{m^{2}}{2}\phi^{2}
\right ] \equiv \int dt d^{3}r L . \label{8}
\end{eqnarray}
Let us rewrite the  classical action (\ref{8}) in the  Bjorken coordinates $(\tau , \eta)$.
Taking into account that  $dt d^{3}r  = d \tau d \sigma$, we get
\begin{eqnarray}
S=\int d \tau d \sigma  L , \label{10}
\end{eqnarray}
where the   Lagrangian density in such coordinates is
\begin{eqnarray}
 L = \frac{1}{2} \left (\frac{\partial \phi}{\partial \tau}\right
 )^{2} -
\frac{1}{2} \frac{1}{\tau^{2}}\left (\frac{\partial \phi}{\partial
\eta}\right )^{2} -  \frac{1}{2}\left (\frac{\partial \phi}{\partial
\textbf{r}_{T}}\right )^{2} - \frac{1}{2}m^{2}\phi^{2}.  \label{11}
\end{eqnarray}
Taking into account that the energy-momentum tensor  is
\begin{eqnarray}
T^{\mu \nu}(x)=\partial^{\mu}\phi\partial^{\nu}\phi -g^{\mu \nu}L,
\label{9}
\end{eqnarray}
it provides that the  local energy density operator in the comoving
frame, $u_{\mu}u_{\nu} T^{\mu\nu}(x)$, is 
\begin{eqnarray}
u_{\mu}u_{\nu} T^{\mu\nu}(x)= \frac{1}{2} \left (\frac{\partial
\phi}{\partial \tau}\right )^{2}+ \frac{1}{2}
\frac{1}{\tau^{2}}\left (\frac{\partial \phi}{\partial \eta}\right
)^{2} +  \frac{1}{2}\left (\frac{\partial \phi}{\partial
\textbf{r}_{T}}\right )^{2}+ \frac{1}{2}m^{2}\phi^{2},  \label{12}
\end{eqnarray}
where we take into account that $u_{\mu}\partial^{\mu}=\partial_{\tau}$. 

It follows from Eq. (\ref{11}) that conjugate momentum $\Pi^{[\tau]}$ with respect
to $\tau$ is
\begin{eqnarray}
\Pi^{[\tau]}=\frac{\partial \phi (x)}{\partial \tau}.
\label{15}
\end{eqnarray}
One can notice from Eqs. (\ref{14}) and (\ref{12})  that
 $H^{[\tau]}$ can be treated as an explicitly $\tau$-dependent ``Hamiltonian''
that generates translations in the timelike direction with respect to $\tau$.
Evidently, such an operator  does not coincide with the Hamiltonian $H$ 
that generates the translation  with respect to $t$. Therefore, defining  with respect 
to the  $H^{[\tau]}$  instantaneous ground state   $|0_{\tau} \rangle$, where $\tau$ is 
the instant to which it refers,  does not coincide with the  global vacuum state in flat Minkowski spacetime,
$|0 \rangle$,  defined with respect to $H$.

In what follows we perform instantaneous diagonalization of $H^{[\tau]}$ in terms of some  appropriate quasiparticle
creation and annihilation operators. It allows us to distinguish contributions of  the corresponding quasiparticles 
 and the  ground state  (quasiparticle vacuum)  to expectation values of relevant  quantities.  
With this aim, it is necessary to find the representation of the canonical commutation
relations  at hypersurface $\tau =
\mbox{const}$,
\begin{eqnarray}
[\phi(x), \Pi^{[\tau]}(x')] = i \frac{1}{\tau}\delta
(\eta- \eta ') \delta^{(2)}(\textbf{r}_{T}-\textbf{r}_{T}'),
\label{16}
\end{eqnarray}
that diagonalizes  $H^{[\tau]}$. 

We start by noting    that  $\phi (x)$ obeys the
 Klein-Gordon equation,
\begin{eqnarray}
(\square - m^{2})\phi (x) =0,  \label{17}
\end{eqnarray}
where $\square = - \partial_{\mu}\partial^{\mu}$ is the d'Alembert
operator associated with the Minkowski spacetime.
It is well known that the  solution of this equation 
in the future light cone can be written with the Hankel functions;
see, e.g., Refs. \cite{Book-1,Akkelin,Hankel-1,Hankel-2,Hankel-3}. 
Then 
\begin{eqnarray}
\phi(x)  = \int_{-\infty}^{+\infty}\frac{d^{2}p_{T}d\mu
}{4\pi\sqrt{2}}[- i e^{\mu \pi /2 +i\mu\eta
+i\textbf{p}_{T}\textbf{r}_{T}}H^{(2)}_{i\mu}(m_{T}\tau)b(\textbf{p}_{T},\mu)+
\nonumber \\
 ie^{- \mu \pi /2 - i\mu\eta -
i\textbf{p}_{T}\textbf{r}_{T}}H^{(1)}_{i\mu}(m_{T}\tau)b^{\dag}(\textbf{p}_{T},\mu)
], \label{18}
\end{eqnarray}
where 
$H^{(1)}_{i\mu}(m_{T}\tau)$  and $H^{(2)}_{i\mu}(m_{T}\tau)$
are the Hankel functions \cite{Hankel},
\begin{eqnarray}
H^{(1)}_{i\mu}(m_{T}\tau) = \frac{1}{i \pi} e^{\mu\pi
/2}\int_{-\infty}^{+\infty}d\vartheta e^{ im_{T} \tau
\cosh\vartheta  -i\mu \vartheta}, \label{19} \\
H^{(2)}_{i\mu}(m_{T}\tau)= - \frac{1}{i \pi}e^{-\mu\pi
/2}\int_{-\infty}^{+\infty}d\vartheta e^{-im_{T} \tau \cosh\vartheta +i\mu \vartheta}. \label{20}
\end{eqnarray}
Here  
\begin{eqnarray}
m_{T}= \sqrt{\textbf{p}_{T}^{2} + m^{2}} \label{21}
\end{eqnarray}
is the so-called transverse mass   and  $\textbf{p}_{T}=(p_{x},p_{y})$ is the transverse momentum.
It is convenient to introduce notations 
\begin{eqnarray}
\Tilde{H}^{(1)}_{i\mu}(x)= H^{(1)}_{i\mu}(x)e^{-\pi \mu/2}, \label{21.1} \\
\Tilde{H}^{(2)}_{i\mu}(x) = H^{(2)}_{i\mu}(x)e^{\pi \mu/2}. \label{21.2}
\end{eqnarray}
Accounting for properties of
the Hankel functions one can see that 
\begin{eqnarray}
\left [\Tilde{H}^{(1)}_{i\mu}(x)\right]^{*}= \Tilde{H}^{(2)}_{i\mu}(x). \label{21.3} 
\end{eqnarray}
Then, assuming 
\begin{eqnarray}
[b(\textbf{p}_{T},\mu), b^{\dag}(\textbf{p}'_{T},\mu ')] =\delta
(\mu- \mu ') \delta^{(2)}(\textbf{p}_{T}-\textbf{p}_{T}'),
\label{22}
\end{eqnarray}
with all other commutators vanishing, and 
by use of the identities (the Wronskian condition)
\begin{eqnarray}
\Tilde{H}^{(2)*}_{i\mu}(x) \overleftrightarrow{\partial_{x}}\Tilde{H}^{(2)}_{i\mu}(x)   = -\frac{4i}{\pi x}, \label{22.1}
\end{eqnarray}
for the Hankel functions and their derivatives,  one  can see  that representation (\ref{18})
realizes the quantization procedure on the hypersurface $\tau =
\mbox{const}$; see Eqs. (\ref{15}) and (\ref{16}). 

It is well known that the vacuum defined with respect 
to the Hankel functions  coincides with the
ordinary Minkowski vacuum defined with respect to the
plane-wave modes,  $b(\textbf{p}_{T},\mu)|0 \rangle =0$ (see, e.g., Ref. \cite{Book-1}).
To argue it, one can  relate plane-wave modes with the Hankel functions. For this aim it is convenient 
to write the solution of the Klein-Gordon equation (\ref{17})  with the plane-wave modes, 
\begin{eqnarray}
\phi(x)  =
\int\frac{d^{3}p}{\sqrt{2\omega_{p}}}\frac{1}{(2\pi)^{3/2}}\left
(e^{-i\omega_{p}t+i \textbf{p} \textbf{r}} a (\textbf{p})+
e^{i\omega_{p}t-i \textbf{p} \textbf{r}} a^{\dag}(\textbf{p})\right
), \label{22.2}
\end{eqnarray}
where
\begin{eqnarray}
\omega_{p} =  \sqrt{\textbf{p}^{2} + m^{2}}.
\label{22.3}
\end{eqnarray}
 The conjugated field
momentum at the hypersurface $t=\mbox{const}$ is
$\Pi=\frac{\partial \phi}{\partial t}$.  The quantization
prescription at such a hypersurface,
\begin{eqnarray}
[\phi(x), \Pi(x')] = i \delta^{(3)}(\textbf{r}-\textbf{r}'),
\label{22.4}
\end{eqnarray}
means that  functions $a^{\dag}(\textbf{p})$ and $a(\textbf{p})$
become creation and annihilation operators, respectively, which
satisfy the following canonical commutation relations:
\begin{eqnarray}
[a(\textbf{p}), a^{\dag}(\textbf{p}')] =
\delta^{(3)}(\textbf{p}-\textbf{p}'), \label{22.5}
\end{eqnarray}
and $[a(\textbf{p}), a(\textbf{p}'),]=[a^{\dag}(\textbf{p}),
a^{\dag}(\textbf{p}')]=0$.
Then, comparing Eq. (\ref{18}) with Eq. (\ref{22.2}) and using  Eqs. (\ref{19}) and (\ref{20}),
one can easily  get (see, e.g., Ref. \cite{Akkelin}) that 
\begin{eqnarray}
a(\textbf{p})= \frac{1}{\sqrt{2\pi \omega_{p}}}\int_{-\infty}^{+\infty}d\mu e^{i\mu\theta}b(\textbf{p}_{T},\mu), \label{22.6} \\
a^{\dag}(\textbf{p})= \frac{1}{\sqrt{2\pi \omega_{p}}}\int_{-\infty}^{+\infty}d\mu e^{-i\mu\theta}b^{\dag}(\textbf{p}_{T},\mu). \label{22.7} 
\end{eqnarray}
Here $\theta$ is momentum rapidity and  $\tanh{\theta}=p_{z}/\omega_{p}$,  $p_{z}$ is the longitudinal  momentum. 
Taking into account Eq. (\ref{22.3}) one can write
\begin{eqnarray}
\omega_{p}= m_{T} \cosh{ \theta}, \label{22.8} \\
p_{z}= m_{T} \sinh { \theta} ,  \label{22.9} 
\end{eqnarray}
where transverse mass $m_{T}$ is defined by Eq. (\ref{21}). It follows from Eqs.  (\ref{22.6}) and (\ref{22.7}) that
``$a$'' and ``$b$'' particles are defined with respect to the same vacuum. 

Substituting (\ref{18}) into Eq. (\ref{12}),   and performing  integrations
over spacetime variables in Eq. (\ref{14}),  we bring $H^{[\tau]}$  to the  form 
\begin{eqnarray}
H^{[\tau]}= \frac{1}{2}\int_{-\infty}^{+\infty}d^{2}p_{T}d\mu \omega (p_{T},\mu,\tau)
[E (p_{T},\mu, \tau) (b^{\dag}(\textbf{p}_{T},\mu)b(\textbf{p}_{T},\mu)+b(\textbf{p}_{T},\mu)b^{\dag}(\textbf{p}_{T},\mu))-
 \nonumber \\ F (p_{T},\mu, \tau) b(\textbf{p}_{T},\mu)b(-\textbf{p}_{T},- \mu)- F^{*}(p_{T},\mu, \tau)
b^{\dag}(\textbf{p}_{T},\mu)b^{\dag}(-\textbf{p}_{T},- \mu)],
\label{23}
\end{eqnarray}
where 
\begin{eqnarray}
\omega (p_{T},\mu, \tau)=\sqrt{m_{T}^{2}+\frac{\mu^{2}}{\tau^{2}}},
\label{24}
\end{eqnarray}
and we introduced notations
\begin{eqnarray}
E (p_{T},\mu,\tau)= \frac{\pi\tau}{4\omega (p_{T},\mu, \tau) } 
\left [ |\partial_{\tau} \Tilde{H}^{(2)}_{i\mu}(m_{T}\tau) |^{2}+\omega^{2} (p_{T},\mu, \tau)| \Tilde{H}^{(2)}_{i\mu}(m_{T}\tau) |^{2}\right ],
\label{25} \\
F (p_{T},\mu,\tau)=\frac{\pi\tau}{4\omega (p_{T},\mu , \tau) } 
\left [ (\partial_{\tau} \Tilde{H}^{(2)}_{i\mu}(m_{T}\tau) )^{2}+\omega^{2} (p_{T},\mu , \tau)( \Tilde{H}^{(2)}_{i\mu}(m_{T}\tau) )^{2}\right ]. \label{26}
\end{eqnarray}
Using Eq. (\ref{22.1}) one can get that 
\begin{eqnarray}
E^{2} (p_{T},\mu,\tau) - 
|F (p_{T},\mu,\tau)|^{2}=1 . \label{27}
\end{eqnarray}

As one sees  from Eq. (\ref{23}), $H^{[\tau]}$ is nondiagonal in the creation and
annihilation operators.  Diagonalization of 
$H^{[\tau]}$ can be performed   by means of 
a canonical  Bogolyubov transformation.\footnote{Our treatment  is  similar to the one required to
diagonalize a Hamiltonian  in an expanding curved spacetime; see, e.g., Ref. \cite{Book-3} and references therein.}
The corresponding  quasiparticle 
 creation, $\xi^{\dag}$,  and destruction, $\xi$,   operators  
 with canonical commutation relations
\begin{eqnarray}
[\xi(\textbf{p}_{T},\mu,\tau),\xi^{\dag}(\textbf{p}'_{T},\mu',\tau)]=
\delta(\mu - \mu') \delta^{(2)}(\textbf{p}_{T}-\textbf{p}'_{T}),
\label{28}
\end{eqnarray}
and
$[\xi^{\dag}(\textbf{p}_{T},\mu,\tau),\xi^{\dag}(\textbf{p}'_{T},\mu',\tau)]=0$, 
$[\xi(\textbf{p}_{T},\mu,\tau),\xi(\textbf{p}'_{T},\mu',\tau)]=0$,
 are related to $b^{\dag}$ and $b$ operators through a Bogolyubov transformation with
$\tau$-dependent  coefficients $\alpha(\textbf{p}_{T},\mu, \tau)$ and $\beta(\textbf{p}_{T},\mu, \tau)$:
\begin{eqnarray}
b(\textbf{p}_{T},\mu)= \alpha(\textbf{p}_{T},\mu, \tau) \xi (\textbf{p}_{T},\mu, \tau) + \beta^{*}(\textbf{p}_{T},\mu, \tau)
\xi^{\dag} (-\textbf{p}_{T},- \mu, \tau), \label{29} \\
b^{\dag}(\textbf{p}_{T},\mu)= \alpha^{*}(\textbf{p}_{T},\mu, \tau) \xi^{\dag} (\textbf{p}_{T},\mu, \tau) + \beta(\textbf{p}_{T},\mu, \tau)
\xi (-\textbf{p}_{T},- \mu, \tau), \label{30} \\
|\alpha(\textbf{p}_{T},\mu, \tau)|^{2}-|\beta(\textbf{p}_{T},\mu, \tau)|^{2}=1. \label{31}
\end{eqnarray}
It follows from  Eqs.  (\ref{29}),  (\ref{30}), and  (\ref{31})  that 
\begin{eqnarray}
\xi (\textbf{p}_{T}, \mu, \tau)= \alpha^{*}(\textbf{p}_{T},\mu, \tau) 
b (\textbf{p}_{T},\mu) - \beta^{*}(\textbf{p}_{T},\mu, \tau)
b^{\dag} (-\textbf{p}_{T},- \mu), \label{31.1} \\
\xi^{\dag} (\textbf{p}_{T}, \mu, \tau)= \alpha(\textbf{p}_{T},\mu, \tau) b^{\dag}
(\textbf{p}_{T},\mu) - \beta(\textbf{p}_{T},\mu, \tau)
b (-\textbf{p}_{T},- \mu). \label{31.2}
\end{eqnarray}
Substituting Eqs. (\ref{29}) and (\ref{30}) into Eq. (\ref{23}) and requiring
diagonalization of $H^{[\tau]}$ in operators 
$\xi^{\dag}$ and $\xi$ we obtain 
\begin{eqnarray}
2 E \alpha\beta -F \alpha^{2}-F^{*}\beta^{2}=0.
 \label{32}
\end{eqnarray}
One can see from  the above equation that $\beta / \alpha $ is a solution of the quadratic equation. Then,
using Eq. (\ref{27}) we get $\frac{\beta}{\alpha}=\frac{E \pm 1}{F^{*}}$. Choosing the solution which tends to zero when
$\tau$ tends to infinity,  we get 
\begin{eqnarray}
\frac{\beta}{\alpha}=\frac{E - 1}{F^{*}}.
 \label{33}
\end{eqnarray}
This implies that 
\begin{eqnarray}
\frac{|\beta|^{2}}{1+|\beta|^{2}}=\frac{(E - 1)^{2}}{|F|^{2}},
 \label{34}
\end{eqnarray}
where we used Eq. (\ref{31}). Taking into account (\ref{27}) we finally  get 
\begin{eqnarray}
|\beta|^{2}=\frac{E - 1}{2}.
 \label{35}
\end{eqnarray}
Also, substituting Eq. (\ref{35}) into  Eq. (\ref{33}) we get 
\begin{eqnarray}
\alpha \beta^{*}=\frac{1}{2}F{^{*}}.
 \label{35.1}
\end{eqnarray}
It is noteworthy that Eqs. (\ref{35}) and (\ref{35.1}) allow us to rewrite 
 Eqs. (\ref{25}) and (\ref{26}) in the form 
\begin{eqnarray}
 \Tilde{H}^{(2)}_{i\mu}(m_{T}\tau) = \left ( \frac{2}{\pi\tau\omega (p_{T},\mu,\tau)}\right)^{1/2} 
 (\alpha^{*}(\textbf{p}_{T},\mu, \tau) + \beta(\textbf{p}_{T},\mu, \tau) ),
\label{35.2} \\
 \partial_{\tau} \Tilde{H}^{(2)}_{i\mu}(m_{T}\tau) = -i \left ( \frac{2\omega (p_{T},\mu, \tau)}{\pi\tau}\right)^{1/2} 
 (\alpha^{*}(\textbf{p}_{T},\mu, \tau) - \beta(\textbf{p}_{T},\mu, \tau) ). \label{35.3}
\end{eqnarray}
Expressions for $\Tilde{H}^{(1)}_{i\mu}(m_{T}\tau)$ and $\partial_{\tau} \Tilde{H}^{(1)}_{i\mu}(m_{T}\tau)$ 
follow from Eq. (\ref{21.3}) and Eqs. (\ref{35.2}) and  (\ref{35.3}).

Under this transformation 
$H^{[\tau]}$ takes the form 
\begin{eqnarray}
H^{[\tau]} = \int_{-\infty}^{+\infty}d^{2}p_{T}d\mu \omega (p_{T},\mu,\tau)\xi^{\dag}(\textbf{p}_{T},\mu, \tau)\xi(\textbf{p}_{T},\mu, \tau),
\label{36}
\end{eqnarray}
where we omitted a constant term taking into account
that such a term is canceled in the expression (\ref{13}) for the quasiequilibrium statistical operator. One can see that
$\omega (p_{T},\mu,\tau)$  has the meaning of the energy of the quasiparticle.
A direct consequence of the Bogolyubov transformation  (\ref{29}), (\ref{30})  and (\ref{31}) is that the
notion of a vacuum is not unique for ``$b$'' and ``$\xi$''
particles. Namely, the  ground state of the ``Hamiltonian'' $H^{[\tau]}$ is  a $\tau$-dependent  highly entangled squeezed state
(see e.g. Refs. \cite{state-1,state-2})  of correlated pairs of $b^{\dag}(\textbf{p}_{T},\mu)$
and  $b^{\dag}(-\textbf{p}_{T},-\mu)$ quanta with zero total momentum.
One can draw an analogy 
between the absence of a unique ground (``vacuum'')  state in the 
Little Bang  created in a relativistic  collision, 
and  absence of a unique vacuum state in  the Big Bang 
cosmological expansion (see, e.g., Refs. \cite{Book-1,Book-3}).

It is now a simple matter to write   
expectation values of $\xi$ and $\xi^{\dag}$  quasiparticle  operators
with  the  quasiequilibrium statistical operator 
(\ref{13}), $\langle  ...\rangle_{\tau}^{q} \equiv Tr[\rho_{\tau}^{q}  ...]$. Because $H^{[\tau]}$ 
has 
the  thermal-like diagonal form  in the quasiparticle representation [see Eq. (\ref{36})] it can be
done utilizing  the thermal Wick theorem 
\cite{Wick} (see also Refs. \cite{Bog,Groot} and Ref. \cite{Akkelin}).
We then obtain
\begin{eqnarray}
\langle \xi^{\dag}\xi^{\dag}\rangle _{\tau}^{q}= \langle
\xi\xi\rangle_{\tau}^{q} = \langle \xi^{\dag}\rangle_{\tau}^{q} = \langle \xi \rangle_{\tau}^{q}= 0, \label{36.1}
\end{eqnarray}
 and 
\begin{eqnarray}
\langle
\xi^{\dag}(\textbf{p}_{T1},\mu_{1},\tau)\xi(\textbf{p}_{T2},\mu_{2},\tau)\rangle_{\tau}^{q}
= \delta(\mu_{1} - \mu_{2})
\delta^{(2)}(\textbf{p}_{T1}-\textbf{p}_{T2})\frac{1}{e^{\frac{
\omega (p_{T1},\mu_{1}, \tau)}{T(\tau)}}-1}. \label{36.2}
\end{eqnarray}
Other expectation values with  $\xi$ and $\xi^{\dag}$ operators   can be calculated utilizing Eqs.
(\ref{36.1}) and  (\ref{36.2}) and the thermal Wick theorem. 
To completely specify the quasiparticle representation, 
one can write $\phi(x)$
and $\partial_{\tau}\phi(x)$ in the following form:
\begin{eqnarray}
\phi(x) =  \frac{1}{4\pi}\frac{1}{\sqrt{\pi \tau}} \int_{-\infty}^{+\infty}
d\mu d^{2}p_{T} \frac{1}{\sqrt{\omega(p_{T},\mu,\tau)}} \times \nonumber \\
\left (- i   e^{i\mu\eta +i\textbf{p}_{T}\textbf{r}_{T}}
  \xi(\textbf{p}_{T},\mu, \tau)+
 i   e^{-i\mu\eta -i\textbf{p}_{T}\textbf{r}_{T}}
  \xi^{\dag}(\textbf{p}_{T},\mu, \tau)\right ), 
\label{36.3} \\
\partial_{\tau}\phi(x) = -  \frac{1}{4\pi}\frac{1}{\sqrt{\pi \tau}} \int_{-\infty}^{+\infty}
d\mu d^{2}p_{T} \sqrt{\omega(p_{T},\mu,\tau)} \times \nonumber \\ 
\left (  e^{i\mu\eta +i\textbf{p}_{T}\textbf{r}_{T}}
  \xi(\textbf{p}_{T},\mu, \tau)
+   e^{-i\mu\eta -i\textbf{p}_{T}\textbf{r}_{T}} 
  \xi^{\dag}(\textbf{p}_{T},\mu, \tau) \right ).
\label{36.4}
\end{eqnarray}
In obtaining the above expressions, we have used  Eqs. (\ref{18}), 
(\ref{21.2}),  (\ref{21.3}), (\ref{31.1}),  (\ref{31.2}), (\ref{35.2}), and (\ref{35.3}).

\section{Momentum spectra of created  particles }

In this section, we  consider  particle production associated with the sudden decoupling of the 
quasiequilibrium state at some hypersurface  $\tau=\tau_{f}$.  We utilize the Heisenberg representation
to describe evolution of the system at $\tau>\tau_{f}$. In this representation the state is time
independent, and because  we have disregarded
self-interactions  of the scalar field we just  assume that   evolution at $\tau>\tau_{f}$ 
is governed by the Klein-Gordon equation.
Note that the  mean 
number of  produced particles in the model can diverge  just because   
both longitudinal and transverse sizes are assumed to be infinite.
Therefore, in order to relate this model to the real world, we  
assume that the effective 
transverse  size of the expanding system is finite but large enough. 
We cannot 
proceed in the same  way with the longitudinal dimension, because  the  boost invariance
of   the model will be then destroyed. Nevertheless,  this difficulty can be circumvented 
if do not integrate particle momentum spectra over $\theta$ and consider 
particle momentum spectra  in  the central rapidity region only.

 To obtain
 meaningful results one needs to  define the normalization condition,
 and  such a condition  should be consistent with the definition of the  $u^{\nu}(x)/T(x)$ field
 at the hypersurface $\tau = \tau_{f}$ by means of Eq. (\ref{1.2}).
 Perhaps the simplest idea is to assume    that
the renormalized energy-momentum tensor is defined   by the  subtraction 
of the expectation value with the   corresponding zero-temperature  ground
state, $T^{\mu \nu}_{ren}(x)= T^{\mu \nu}(x) -\langle 0_{\tau_{f}}| T^{\mu \nu}(x)| 0_{\tau_{f}} \rangle$;
see Ref. \cite{Tinti}.
Such a subtraction can be understood as  the normal ordering with respect
to the $\xi^{\dag}(\textbf{p}_{T},\mu,\tau_{f})$ and $\xi(\textbf{p}_{T},\mu,\tau_{f})$ operators.
However, while such a renormalization procedure has some attractive features,\footnote{It was shown in Ref. \cite{Tinti}
that the expectation value of the energy-momentum tensor of the quasiparticles,
$\langle T^{\mu \nu}(\tau,\textbf{r}_{T},\eta)\rangle _{\tau}^{q}-
\langle 0_{\tau}| T^{\mu \nu}(\tau,\textbf{r}_{T},\eta)| 0_{\tau} \rangle$, has the perfect fluid form.}  it 
has a drawback from a physical perspective. Namely,    the defining property,
$\langle 0_{\tau_{f}}| T^{\mu \nu}_{ren}(x)| 0_{\tau_{f}} \rangle =0$,
is tailored to the proper time $\tau_f$ and to the quasiparticle vacuum $| 0_{\tau_{f}} \rangle$.  Then
 the renormalized energy-momentum tensor does not vanish 
 in the ordinary Minkowski vacuum.
 Therefore,  we define the renormalization procedure by 
subtracting the 
expectation value with the  Minkowski vacuum, 
\begin{eqnarray}
T^{\mu \nu}_{ren}(x)= T^{\mu \nu}(x) -\langle 0| T^{\mu \nu}(x)| 0 \rangle . \label{37}
\end{eqnarray}

We assume that field operators at  $\tau=\tau_{f}$ are defined in the quasiparticle representation   by Eqs. (\ref{36.3})
and (\ref{36.4}),
and that their further evolution is governed by the Klein-Gordon equation. 
To perform a smooth interpolation  between the quasiparticle 
vacuum at $\tau=\tau_{f}$,  $| 0_{\tau_{f}} \rangle$, and the ordinary Minkowski vacuum, $| 0 \rangle$, we take into 
account that the Bogolyubov coefficient $\beta$ [see Eqs. (\ref{25}) and (\ref{35})] tends to zero  when
$\tau$ tends to infinity. It allows us to use the  quasiparticle creation, 
$\xi^{\dag}(\textbf{p}_{T},\mu,\tau)$, and annihilation,
$\xi(\textbf{p}_{T},\mu,\tau)$, operators at intermediate proper times $\tau_{f}<\tau < \infty$ to 
interpolate between quasiparticle 
and particle degrees of freedom.
Note that  the instantaneous quasiparticle  vacuum, $| 0_{\tau} \rangle$,  is situated 
in the future light cone  and 
is defined as  the
eigenstate of the normal ordered with respect to the $\xi^{\dag}(\textbf{p}_{T},\mu,\tau)$ 
and $\xi(\textbf{p}_{T},\mu,\tau)$  operator $H^{[\tau]}=\int  d\sigma u_{\nu}(x)u_{\mu}(x)T^{\mu \nu}(x)$ with the
lowest eigenvalue
at  the three-dimensional  hypersurface $\tau = \mbox{const}$.  Since $H^{[\tau]}$ turns out to be $\tau$-dependent, 
the corresponding  instantaneous vacuum state  continuously evolves
with $\tau$, and  approaches the Minkowski vacuum state at 
asymptotic proper times: $| 0_{\tau \rightarrow \infty} \rangle = | 0 \rangle $.

It is worth noting that $\xi^{\dag}(\textbf{p}_{T},\mu,\tau)$ and 
$\xi(\textbf{p}_{T},\mu,\tau)$ have  nontrivial time dependence
even while the field satisfies  the free evolution equation. 
It leads to nonconservation of the quasiparticle number and momentum spectra during evolution  along the 
timelike direction $u^{\mu}(x)$.
To determine the spectrum of quasiparticles at intermediate proper times,  
one can utilize Eqs. (\ref{31.1}) and (\ref{31.2}) and relate 
$ \xi^{\dag}(\textbf{p}_{T},\mu,\tau)$ and 
$\xi(\textbf{p}_{T},\mu,\tau)$ with $b^{\dag}(\textbf{p}_{T},\mu)$ 
and $b(\textbf{p}_{T},\mu)$, and then utilize 
Eqs. (\ref{29}) and  (\ref{30}) to relate 
$b^{\dag}(\textbf{p}_{T},\mu)$ 
and $b(\textbf{p}_{T},\mu)$ with 
$ \xi^{\dag}(\textbf{p}_{T},\mu,\tau_{f})$ and 
$\xi(\textbf{p}_{T},\mu,\tau_{f})$.
Then the time-dependent creation, $\xi^{\dag}(\textbf{p}_{T},\mu, \tau)$, and annihilation,
$\xi(\textbf{p}_{T},\mu, \tau)$,   operators    are
related to   $\xi(\textbf{p}_{T},\mu, \tau_{f})$  and $\xi(\textbf{p}_{T},\mu, \tau_{f})$ operators via the
time-dependent Bogolyubov transformation,
\begin{eqnarray}
\xi (\textbf{p}_{T}, \mu, \tau)= u(\textbf{p}_{T},\mu, \tau) 
\xi (\textbf{p}_{T},\mu, \tau_{f}) + v^{*}(\textbf{p}_{T},\mu, \tau)
\xi^{\dag} (-\textbf{p}_{T},- \mu, \tau_{f}), \label{40} \\
\xi^{\dag} (\textbf{p}_{T}, \mu, \tau)= u^{*}(\textbf{p}_{T},\mu, \tau) 
\xi^{\dag} (\textbf{p}_{T},\mu, \tau_{f})  + v(\textbf{p}_{T},\mu, \tau)
\xi (-\textbf{p}_{T},- \mu, \tau_{f}), \label{41}
\end{eqnarray}
where the Bogolyubov coefficients $u(\textbf{p}_{T},\mu, \tau)$ and $v(\textbf{p}_{T},\mu, \tau)$ are
\begin{eqnarray}
 u(\textbf{p}_{T},\mu, \tau) = \alpha^{*}(\textbf{p}_{T},\mu, \tau)\alpha(\textbf{p}_{T},\mu, \tau_{f})-
\beta^{*}(\textbf{p}_{T},\mu, \tau)\beta(-\textbf{p}_{T},-\mu, \tau_{f}), \label{42} \\
 v(\textbf{p}_{T},\mu, \tau) = \alpha(\textbf{p}_{T},\mu, \tau)\beta(\textbf{p}_{T},\mu, \tau_{f})-
\beta(\textbf{p}_{T},\mu, \tau)\alpha(-\textbf{p}_{T},-\mu, \tau_{f}). \label{43}
\end{eqnarray}
Taking into account symmetry $\textbf{p}_{T} \rightarrow - \textbf{p}_{T} $, $\mu \rightarrow -\mu$ properties 
of the coefficients, it is straightforward
to confirm that
\begin{eqnarray}
|u(\textbf{p}_{T},\mu, \tau)|^{2}-|v(\textbf{p}_{T},\mu, \tau)|^{2} =1.
\label{44}
\end{eqnarray}
Equations (\ref{40}) and   (\ref{41}) mean
that the  quasiparticle vacuum state at $\tau =\tau_{f} $   is the 
excited state  from the perspective of the quasiparticles at $\tau > \tau_{f} $. 
It provides that evolution is accompanied by  quasiparticle production 
in an analogy with particle production in an expanding universe  \cite{Parker} (see also Refs.
\cite{Book-1,Book-3,Book-2,Book-4}.

Taking into account Eqs. (\ref{36.2}), (\ref{40}),   and (\ref{41}),  we get that at $\tau > \tau_{f}$
\begin{eqnarray}
\langle
\xi^{\dag}(\textbf{p}_{T1},\mu_{1},\tau)\xi (\textbf{p}_{T2},\mu_{2},\tau)\rangle_{\tau_{f}}^{q}
= \delta(\mu_{1} - \mu_{2})
\delta^{(2)}(\textbf{p}_{T1}-\textbf{p}_{T2}) \times \nonumber \\ \left (\frac{1+2|v(\textbf{p}_{T1},\mu_{1}, \tau)|^{2}}{e^{\frac{
\omega (p_{T1},\mu_{1}, \tau_{f})}{T(\tau_{f})}}-1}+ |v(\textbf{p}_{T1},\mu_{1}, \tau)|^{2}\right ). \label{47}
\end{eqnarray}
It is worth noting  that   $ \langle \xi^{\dag}(\textbf{p}_{T},\mu,\tau)\rangle_{\tau_{f}}^{q} = \langle \xi (\textbf{p}_{T},\mu,\tau) \rangle_{\tau_{f}}^{q}= 0 $, and that
$\langle \xi^{\dag}(\textbf{p}_{T1},\mu_{1},\tau)\xi^{\dag}(\textbf{p}_{T2},\mu_{2},\tau)\rangle_{\tau_{f}}^{q} $
as well as $\langle \xi(\textbf{p}_{T1},\mu_{1},\tau) \xi (\textbf{p}_{T2},\mu_{2},\tau)\rangle_{\tau_{f}}^{q} $
are not identically zero at $\tau > \tau_{f}$. 
To calculate the  corresponding one-particle momentum spectra  at asymptotic times,
\begin{eqnarray}
\omega_{p}\frac{d^{3}N}{d^{3}p }= \frac{d^{3}N}{d^{2}p_{T}d \theta } = \omega_{p}  \langle a^{\dag}(\textbf{p})a (\textbf{p})\rangle_{\tau_{f}}^{q},
 \label{48}
\end{eqnarray}
  one needs to relate plane-wave modes with quasiparticle modes at $\tau=\tau_{f}$. 
First, taking into account that $\beta$ tends to zero when $\tau$ tends to infinity and using Eqs. 
(\ref{29}) and (\ref{30})
and Eq. (\ref{47}),   we get  
\begin{eqnarray}
 \alpha^{*}(\textbf{p}_{T1},\mu_{1},\tau \rightarrow \infty)\alpha(\textbf{p}_{T2},\mu_{2},\tau \rightarrow \infty ) \langle
\xi^{\dag}(\textbf{p}_{T1},\mu_{1},\tau \rightarrow \infty)\xi (\textbf{p}_{T2},\mu_{2},\tau \rightarrow \infty )\rangle_{\tau_{f}}^{q}=  \nonumber \\
\langle
b^{\dag}(\textbf{p}_{T1},\mu_{1})b (\textbf{p}_{T2},\mu_{2})\rangle_{\tau_{f}}^{q}
=   \delta(\mu_{1} - \mu_{2})
\delta^{(2)}(\textbf{p}_{T1}-\textbf{p}_{T2})\times \nonumber \\  \left (\frac{1+2|\beta(\textbf{p}_{T1},\mu_{1}, \tau_{f})|^{2}}{e^{\frac{
\omega (p_{T1},\mu_{1}, \tau_{f})}{T(\tau_{f})}}-1}+ |\beta(\textbf{p}_{T1},\mu_{1}, \tau_{f})|^{2}\right ). \label{49}
\end{eqnarray}
Here we took into account that $|\alpha(\textbf{p}_{T},\mu,\tau \rightarrow \infty )|=1$.
It is worth noting  that   $ \langle b^{\dag}\rangle_{\tau_{f}}^{q} = \langle b \rangle_{\tau_{f}}^{q}= 0 $, and that
$\langle b^{\dag}b^{\dag}\rangle_{\tau_{f}}^{q} $
as well as $\langle bb \rangle_{\tau_{f}}^{q} $ are not identically zero. 
Then, utilization of Eqs.  (\ref{22.6}) and (\ref{22.7})  yields 
\begin{eqnarray}
\langle
a^{\dag}(\textbf{p}_{1})a (\textbf{p}_{2})\rangle_{\tau_{f}}^{q}
= \frac{1}{2\pi \sqrt{ \omega_{p1}\omega_{p2}}}\delta^{(2)}(\textbf{p}_{T1}-\textbf{p}_{T2}) \times \nonumber \\ \int_{-\infty}^{+\infty}
d\mu e^{-i\mu(\theta_{1} - \theta_{2})}\left (\frac{1+2|\beta(\textbf{p}_{T1},\mu, \tau_{f})|^{2}}{e^{\frac{
\omega (p_{T1},\mu_{1}, \tau_{f})}{T(\tau_{f})}}-1}+ |\beta(\textbf{p}_{T1},\mu, \tau_{f})|^{2}\right ). \label{50}
\end{eqnarray}
Nonthermal contributions into Eqs. (\ref{49}) and  (\ref{50})  appear 
due to the difference between the   quasiparticle  vacuum state and the 
global Minkowski vacuum state: the zero-temperature ground state $ | 0_{\tau_{f}} \rangle$ is an excited 
state with  respect to the observed plane-wave particles.  Approximations of the  corresponding expressions
for $m_{T}\tau_{f} \gg 1$ 
can be found in Ref. \cite{Akkelin}.
In a sense,  Eq. (\ref{50}) immediately follows if one  relates
plane-wave modes with quasiparticle modes 
at $\tau=\tau_{f}$ by means of Eqs. (\ref{22.6}), (\ref{22.7}), (\ref{29}) and (\ref{30}). However,
one needs to take into account that Eq. (\ref{50}) describes the actual
particle momentum spectra at asymptotic times only. 

Utilizing ultraviolet
asymptotic expansions for the Hankel functions one can show   that the 
expectation value of the renormalized energy-momentum tensor (\ref{37}),
$\langle T^{\mu \nu}_{ren}(x)\rangle _{\tau_{f}}^{q}$, diverges. The divergent 
contributions appear due to the difference between the   quasiparticle  vacuum state and the 
global Minkowski vacuum,
$(\langle 0_{\tau_{f}}| T^{\mu \nu}(x)| 0_{\tau_{f}} \rangle -\langle 0| T^{\mu \nu}(x)| 0 \rangle)$, 
see Ref. \cite{Tinti}. 

As a final comment we would like to point out  that  if one neglects  the vacuum particle production 
terms in Eq. (\ref{50}), then  
the one-particle 
momentum spectrum  (\ref{48})  coincides with 
the spectrum   of the  local-equilibrium ideal Bose-Einstein gas.  Indeed,
making  the substitution $\delta^{(2)}(\textbf{p}_{T1}-\textbf{p}_{T2}) \rightarrow (2\pi)^{-2}R_{T}^{2}$ 
at $\textbf{p}_{T1} = \textbf{p}_{T2}$, changing the 
integration variable $\mu = (m_{T}\tau_{f})\sinh(\eta-\theta)$, and calling the
integration over the transverse dimension the effective transverse size of the system $R_{T}^{2}$,  we  obtain 
\begin{eqnarray}
\frac{R_{T}^{2}}{(2\pi)^{3} }\int_{-\infty}^{+\infty}d\eta
m_{T}\tau_{f}\cosh (\eta -\theta )\frac{1}{e^{\frac{m_{T}\cosh
(\eta-\theta)}{T(\tau_{f})}}-1}=
\int_{\sigma_{f}}d\sigma_{f}
u^{\mu}p_{\mu} \frac{1}{(2\pi)^{3}} \frac{1}{e^{\frac{u^{\mu}p_{\mu}}{T(\tau_{f})}}-1} , \label{51}
\end{eqnarray}
where $ (2\pi)^{-3}  (e^{\frac{u^{\mu}p_{\mu}}{T(\tau_{f})}}-1)^{-1} $
corresponds to the Bose-Einstein  local equilibrium distribution
function of the ideal gas, $u^{\mu}$ is given by (\ref{5}),  and $p^{\mu}=(m_{T}\cosh{\theta},\textbf{p}_{T},
m_{T}\sinh{\theta})$.

\section{Conclusions}

Acceptability of hydrodynamics for the description of 
particle production in the relativistic nucleus and particle collisions  means that a true state of the system can be
substituted by some proxy state. Such a reduced description can be 
fulfilled  based on   the  quasiequilibrium
statistical operator. In general,  the ground state of this  statistical operator  may not coincide with the 
Minkowski vacuum. In the present work, we  analyze particle production  after sudden   decoupling  of the  
quasiequilibrium state of the expanding system created in an ultrarelativistic  collision. 
Because the 
spacetime region occupied by the matter produced 
in an ultrarelativistic  collision lies in the future light cone of a  collision,
initial conditions and subsequent  evolution 
are described by utilizing some appropriate curved 
coordinate  system in the light cone. To make the problem tractable,  we consider a simple but reliable model of the
quasiequilibrium state of the  noninteracting boost-invariant expanding scalar quantum field. 
We have performed an exact diagonalization of the model in the quasiparticle representation 
by means of the  Bogolyubov transformation. It allowed  us to explicitly disclose the 
zero-temperature ground state (the instantaneous   quasiparticle vacuum).
Then, we analyze particle production induced by the sudden  decoupling of the quasiequilibrium state of
the system.  It is noteworthy that
the mechanism of particle creation from 
little bang fireballs created in ultrarelativistic heavy ion and particle collisions 
has some similarities  with the cosmological particle creation. 

It is worth noting that contributions to particle momentum spectra from 
the quasiequilibrium  ground state
 could be observed in relativistic particle and nucleus collisions.  Specifically, it was proposed  \cite{Akkelin} that 
peculiarities  of the measured in $p+p$ collisions at the
LHC \cite{Alice-1,Alice-2,Atlas,CMS-1,CMS-2}  Bose-Einstein 
momentum correlations of two identical charged pions could be attributed   
to the two-source  mechanism of particle emission associated with the decoupling of the  quasiequilibrium 
state.\footnote{Notice that the existing of two scales in $p + p$
collisions was also proposed in Ref. \cite{scale} based on a different
underlying physical picture.}   To make possible  quantitative 
 comparison with experimental data, 
a generalization of the
model (accounting for field interactions, expansion in the transverse direction, 
feed-downs from the resonance decays, etc.) should be performed.

%\begin{acknowledgments}

%\end{acknowledgments}


\begin{thebibliography}{99}

\bibitem{Shen} Chun Shen,  Nucl. Phys. A \textbf{1005}, 121788 (2021)   [arXiv:2001.11858].
\bibitem{Zubarev-1} D.N. Zubarev, A.V. Prozorkevich, S.A. Smolyanskii, TMF \textbf{40}, 394
(1979) [Theor.  Math. Phys.  \textbf{40}, 821  (1979)].
\bibitem{Zubarev-2}  A. Hosoya, M.-a. Sakagami, M. Takao, Ann.
Phys. (N.Y.) \textbf{154}, 229 (1984).
\bibitem{Zubarev-3}  D. Zubarev, V. Morozov, G. R\"{o}pke, \textit{Statistical Mechanics of
Nonequilibrium Processes. Volume 1: Basic Concepts. Kinetic Theory}
(Berlin, Akademie Verlag, 1996).
\bibitem{Zubarev-4}  D. Zubarev, V. Morozov, G. R\"{o}pke, \textit{Statistical Mechanics of Nonequilibrium
Processes. Volume 2: Relaxation and Hydrodynamic Processes} (Berlin,
Akademie Verlag, 1997).
\bibitem{Bec-1} F. Becattini, L. Bucciantini, E. Grossi, L. Tinti, Eur. Phys. J.
C \textbf{75}, 191 (2015) [arXiv:1403.6265].
\bibitem{Bec-2} T. Hayata,  Y. Hidaka, T. Noumi, M. Hongo,  Phys. Rev. D \textbf{92}, 065008 (2015)
[arXiv:1503.04535].
\bibitem{Bec-3} A. Harutyunyan, A. Sedrakian, D.H. Rischke,
Particles \textbf{1}, 155 (2018) [arXiv:1804.08267].
\bibitem{Bec-4} F. Becattini, M. Buzzegoli and E. Grossi,  Particles \textbf{2},
197 (2019) [arXiv:1902.01089].
\bibitem{Bec-5} D. Blaschke, G. R\"{o}pke, D.N. Voskresensky,  V.G. Morozov, 
Particles  \textbf{3},  380 (2020) [arXiv:2004.05401].
\bibitem{Janes-1} E.T. Jaynes, Phys. Rev. \textbf{106}, 620 (1957).
\bibitem{Janes-2} E.T. Jaynes, Phys. Rev. \textbf{108}, 171 (1957).
\bibitem{Sinyukov-1} Yu.M. Sinyukov, Preprint ITP-93-8E, 1993.
\bibitem{Sinyukov-2} Yu.M. Sinyukov, Nucl. Phys. A \textbf{566},
589c (1994).
\bibitem{Sinyukov-3} Yu.M. Sinyukov, Heavy Ion Phys. \textbf{10}, 113 (1999) [arXiv:nucl-th/9909018].
\bibitem{Akkelin} S.V. Akkelin, Eur. Phys. J. A  \textbf{55}, 78 (2019) [arXiv:1812.03905].
\bibitem{Bjorken} J.D. Bjorken, Phys. Rev. D \textbf{27}, 140 (1983).
\bibitem{Book-1}  N.D. Birrell and P.C.W. Davies, \textit{Quantum Fields in Curved Space}
(Cambridge University Press, Cambridge, England, 1984).
\bibitem{Hankel-1} C.M. Sommerfield, Ann. Phys. (N.Y.) \textbf{84}, 285
(1974).
\bibitem{Hankel-2} N.B. Narozhny, A.M. Fedotov, B.M. Karnakov, V.D. Mur, V.A.
Belinskii, Phys. Rev. D \textbf{65},  025004 (2001)
[arXiv:hep-th/9906181].
\bibitem{Hankel-3} L.C.B. Crispino, A. Higuchi, G.E.A. Matsas, Rev. Mod. Phys. \textbf{80}, 787 (2008) [arXiv:0710.5373].
\bibitem{Hankel} I.S. Gradshteyn,   I.M. Ryzhik,  \textit{Table of Integrals, Series and Products}
 (Academic, New York, 1980).
 \bibitem{Book-3} A.A. Grib, S.G. Mamayev, and V.M. Mostepanenko, \textit{Vacuum Quantum Effects in Strong Fields}
(Friedmann Laboratory Publishing, St. Petersburg, 1994).
\bibitem{state-1} L. Mandel, E. Wolf, \textit{Optical Coherence and Quantum Optics} 
(Cambridge University Press, Cambridge, England,  1995).
\bibitem{state-2} J.-P. Gazeau, \textit{ Coherent States in Quantum Physics}
(Wiley-VCH Verlag Gmbh $\&$ Co. KGaA, Weinheim, 2009).
\bibitem{Wick} C. Bloch, C. De Dominicis, Nucl. Phys. \textbf{7}, 459
(1958).
\bibitem{Bog} N.N. Bogolubov, N.N. Bogolubov, Jr., \textit{An Introduction to
 Quantum Statistical Mechanics} (Gordon and Breach, New York, 1992).
\bibitem{Groot}   S.R. de Groot, W.A. van Leeuwen, Ch. G. van Weert,
\textit{Relativistic Kinetic Theory} (North-Holland, Amsterdam,
1980).
\bibitem{Tinti} D. Rindori, L. Tinti, F. Becattini, D. Rischke, arXiv:2102.09016.
\bibitem{Parker}  L. Parker, Phys. Rev. \textbf{183}, 1057 (1969).
\bibitem{Book-2}  Stephen A. Fulling. \textit{Aspects of Quantum Field Theory 
in Curved Space-Time} (Cambridge University Press, Cambridge, England,  1989). 
\bibitem{Book-4} L.E. Parker and D.J. Toms, \textit{Quantum Field Theory in Curved Spacetime}
(Cambridge University Press, Cambridge, England, 2009).
\bibitem{Alice-1} K. Aamodt \textit{et al.} (ALICE Collaboration),  Phys. Rev. D \textbf{84}, 112004 (2011).
\bibitem{Alice-2} S. Acharya  \textit{et al.} (ALICE Collaboration), J. High Energy Phys. 09 
(2019) 108 [arXiv:1901.05518]. 
\bibitem{Atlas}  ATLAS Collaboration, Eur. Phys. J. C \textbf{75}, 466 (2015).
\bibitem{CMS-1} A.M. Sirunyan \textit{et al.} (CMS Collaboration), Phys. Rev. C \textbf{97}, 064912
(2018) [arXiv:1712.07198]. 
\bibitem{CMS-2}  A.M. Sirunyan \textit{et al.}  (CMS Collaboration),  J. High Energy Phys. 03 (2020) 014
 [arXiv:1910.08815].  	
\bibitem{scale} V.A. Khoze, A.D. Martin, M.G. Ryskin, V.A. Schegelsky, Eur. Phys. J. C \textbf{76},  193 (2016)
[arXiv:1601.08081].

\end{thebibliography}
\end{document}